\documentclass[aoas,nameyear,seceqn,dvips]{arximspdf}
\usepackage{graphics}
\doi{10.1214/09-AOAS325}
\volume{4}
\issue{3}
\pubyear{2010}
\firstpage{1451}
\lastpage{1475}


\begin{document}
\begin{frontmatter}

\title{A smoothing approach for masking spatial data}
\runtitle{A smoothing approach for masking spatial data}

\begin{aug}
\author[A]{\fnms{Yijie} \snm{Zhou}\corref{}\ead[label=e1]{yijie\_zhou@merck.com}},
\author[B]{\fnms{Francesca} \snm{Dominici}\thanksref{a1}\ead[label=e2]{fdominic@hsph.harvard.edu}} \and
\author[C]{\fnms{Thomas A.} \snm{Louis}\thanksref{a2}\ead[label=e3]{tlouis@jhsph.edu}}
\runauthor{Y. Zhou, F. Dominici and T. A. Louis}
\affiliation{Merck Research Laboratories, Harvard University and Johns
Hopkins University}
\address[A]{Y. Zhou\\ Merck Research Laboratories\\ P.O. Box 2000\\
RY34-A304\\
Rahway, New Jersey 07065\\USA\\\printead{e1}} %adresu isvedimo
%komanda gale!
\address[B]{F. Dominici\\ Department of
Biostatistics\\ Harvard University\\ 655 Huntington Avenue\\
Boston, Massachusetts 02115\\USA\\ \printead{e2}}
\address[C]{T. A. Louis\\
Department of Biostatistics\\ Johns Hopkins University\\ 615
N. Wolfe St.\\ Baltimore, Maryland 21205\\USA\\ \printead{e3}}
\thankstext{a1}{Supported in part by the National Institute for Environmental
Health Sciences Grant R01ES012054 and by the Environmental
Protection Agency Grants R83622 and RD83241701.
The content is solely the responsibility of the authors
and does not necessarily represent the official views of the National
Institute of Environmental Health Science nor of the Environmental Protection
Agency.}
\thankstext{a2}{Supported in part by the National Institute of
Diabetes, Digestive and Kidney
Diseases Grant R01 DK061662.
The content is solely the
responsibility of the authors and does not necessarily represent the official
views of the National Institute of Diabetes, Digestive and Kidney Diseases.}
\end{aug}

% HISTORY:
\received{\smonth{10} \syear{2007}}
\revised{\smonth{12} \syear{2009}}

% ABSTRACT
%
\begin{abstract}
Individual-level health data are often not publicly available due
to confidentiality; masked data are released instead. Therefore,
it is important to evaluate the utility of using the masked data
in statistical analyses such as regression. In this paper we
propose a data masking method which is based on spatial smoothing
techniques. The~proposed method allows for selecting both the form
and the degree of masking, thus resulting in a large degree of
flexibility. We investigate the utility of the masked data sets in
terms of the mean square error~(MSE) of regression parameter
estimates when fitting a Generalized Linear Model~(GLM) to the
masked data. We also show that incorporating prior knowledge on
the spatial pattern of the exposure into the data masking may
reduce the bias and MSE of the parameter estimates. By evaluating
both utility and disclosure risk as functions of the form and the
degree of masking, our method produces a risk-utility profile
which can facilitate the selection of masking parameters. We apply
the method to a study of racial disparities in mortality rates
using data on more than 4 million Medicare enrollees residing in
2095 zip codes in the Northeast region of the United States.
\end{abstract}

% KEYWORDS
%
\begin{keyword}
\kwd{Statistical disclosure limitation}
\kwd{data masking}
\kwd{data utility}
\kwd{disclosure risk}
\kwd{spatial smoothing}.
\end{keyword}

\end{frontmatter}
%

%s1 ###
\section{Introduction}\label{introchap4}
Individual-level information such as health data collected by, for
example, government agencies, are often not publicly available in
order to preserve confidentiality. On the other hand, there is
public demand on these individual-level data for research
purposes. As an example, associations of individual health with
various risk factors are of great interest and concern nowadays.
Statistical research that addresses these two competing needs is
known as \textit{statistical disclosure limitation}, where a large
number of methods are developed on how to process and release
information that is subject to confidentiality
concern~[\citet{DuncLambdisc1986}; \citet{FienWillintr1998}; \citeauthor{WillWaalstat1996}
(\citeyear{WillWaalstat1996}, \citeyear{WillWaalelem2001})].
In this paper we refer to those methods that alter the original
data values as ``data masking.'' Corresponding to the two competing
needs, a~data masking method should be evaluated from both the
utility of the masked data which represents the information
retained after the masking, and the disclosure risk of the masked
data which is the risk that a data intruder can obtain
confidential information (e.g., obtain original data values and/or
identify an individual to whom a data record belongs). Ideally,
masked data would have low disclosure risk while preserving data
utility as much as possible.

Examples of commonly used data masking methods include aggregated
tabular counts for categorical data~[\citet{FienSlavmaki2004}],
data swapping which exchanges values between selected records,
with its various
extensions~[\citet{DaleReisdata1982}; \citet{FienMcIndata2005}], cell
suppression where certain cells of contingency tables are not
displayed~[\citet{Coxnetw1995}], simulating synthetic data which
have the same (conditional) distribution as the original
data~[\citet{Rubicomm1993}; \citet{FienMakoSteedisc1998};
\citet{RaghReitRubimult2003}; \citeauthor{Reitinfe2003} (\citeyear{Reitinfe2003}, \citeyear{Reitrele2005})],
and additive random noise for continuous
variables~[\citet{Kimmeth1986}; \citet{SullFulluse1989}; \citet{Fullmask1993}; \citet{trotfienmarkovmeyer2004}],
etc.

Among these methods, data aggregation, data swapping, additive
random noise and many other methods can be formulated as {\it
matrix masking} [\citet{DuncPearrepl1991}]. Suppose data on $n$
observations and $p$ variables are stored in a $n\times p$ matrix.
Matrix masking takes the general form of $\mathbf{Z}^\ast=A\mathbf{Z}B+C$,
where $\mathbf{Z}$ is the original data matrix and $\mathbf{Z}^\ast$
is the
masked data matrix. Matrices $A$, $B$ and $C$ are
row~(observation) operator, column~(variable) operator and random
noise, respectively. Links between the above masking methods to
matrix masking are investigated in~\citet{DuncPearrepl1991},
\citet{Coxmatr1994}, \citet{Fienconf1994} and
\citet{FienMakoSteedisc1998}.

Measuring and evaluating utility of masked data is important. In
general there are two classes of utility measures. One is global
utility measures which reflect the general distribution of masked
data compared to that of the original data and are not specific to
any analysis. Such measures include the number of swaps in data
swapping, the added variance in the additive random noise
approach, differences between continuous original and masked data
in their first and second moments, etc. More sophisticated
measures that compare distributions of masked and original data
can be found in \citet{DobraKarrSanilFien2002},
\citet{GomaKarrSanidata2005} and
\citet{WooReitOganKarglob2009}. In addition, Bayesian
decision theory-based utility is discussed in~\citet{trotfien2002}
and~\citet{DobrFienTrotasse2003}.

The~second class of utility measures is analysis-specific tailored
to analysts' inference. For the utility associated with regression
inference, \citet{KarKohnOganReitSanifram2006} examine the
overlap in the confidence intervals of linear regression
coefficients estimated with original and masked data.
\citet{Kimmeth1986} and \citet{Fullmask1993} show for the
additive random noise approach that if masked data preserve the
first two moments of original data, then coefficient estimates
from linear regression using masked data are (approximately)
unbiased. In addition, the methods of aggregated tabular counts
and data swapping can produce valid results for loglinear models
because they preserve the marginal total of contingency tables.
This is equivalent to preserving sufficient statistics for
loglinear models, given that the margins of all higher-order
interactions that appear in the model are
preserved~[\citet{FienSlavmaki2004}; \citet{FienMcIndata2005}].
Recently, \citet{SlavLee2009} investigated logistic regression
inference for contingency tables that preserve marginal total or
conditional probabilities. However, for a general data structure
additional research is needed. For example, bias and variance of
parameter estimates from nonlinear regression using masked data
are not quantified as functions of masking parameters.

We propose a special case of matrix masking where we construct
row (observation) transformed data, that is, $\mathbf{Z}^\ast
=A\mathbf{Z}$, using
spatial smoothing. We investigate the mean square error (MSE) of
the regression parameter estimates when fitting a Generalized
Linear Model~(GLM) to the masked data, and we provide guidance on
how to select the masking parameters to reduce the MSE.
Specifically, for both regressors and outcome we construct masked
data which are weighted averages of the original individual-level
data by using linear smoothers. The~shape of the smoothing weight
function defines the ``form'' of masking and the smoothness
parameter measures the ``degree'' of masking. By choosing an
appropriate weight function and smoothness parameter value, the
masked data can account for prior knowledge on the spatial pattern
of individual-level data, and parameter estimates from nonlinear
regression using such masked data may be less subject to bias and
MSE. Although data utility is our main focus, we also evaluate
identification disclosure risk. We consider the scenario wherein a
data intruder has correct information on the risk factor
regressors (e.g., exposure or demographic data) from some external
data sources, and his/her objective is to obtain the confidential
information on the health outcome through record matching. Using
our method, we can evaluate both the utility and the disclosure
risk as functions of the form and the degree of masking, which
produces a risk-utility profile and can facilitate the selection
of the masking parameters. We also derive a closed-form expression
for calculating the first-order bias of the regression parameter
estimates when estimated using the masked data, for any assumed
distribution of the outcome given the regressors in the
exponential family.

We apply our method to a study of racial disparities in risks of
mortality for a large sample of the U.S. Medicare population. This
study consists of more than 4~million individuals in the Northeast
region of the United States. We develop and apply statistical
models to estimate the age and gender adjusted association between
race and risks of mortality when using both the original
individual-level data and the masked data. The~estimated
association obtained from using the original individual-level data
is the gold-standard, and we compare it to the estimated
association obtained from using the masked data. We also calculate
the identification disclosure risk of the masked data sets.

In Section~\ref{sec2} we detail the method, and in Section~\ref{sec3} we present the
simulation studies. In Section~\ref{sec4} we apply our method to the
Medicare data set, and in Section~\ref{sec5} we discuss the method and the
results. The~R code is provided in the Supplement [Zhou, Dominici and Louis (\citeyear{ZhDoLo2010b})], while the
Medicare data set is not provided due to a confidentiality agreement.
Derivation of the closed-form expression for the first-order bias
of the GLM regression parameter estimates when estimated using the
masked data is presented in the \hyperref[append]{Appendix}.

%s2 ###
\section{Methods}\label{sec2}

%s2.1 ###
\subsection{Matrix masking using spatial smoothing}\label{method}

Assume that the outcome variable $Y$ and the regressors $\mathbf{X}$ are
spatial processes $\{Y(s),\mathbf{X}(s)\}$, and the observed
individual-level data $\{(Y_i,\mathbf{X}_i),i=1,\ldots,N \}$ are
realizations of the spatial processes at locations
$\mathbf{s}=\{s_1,\ldots, s_N\}$, that is, $\mathbf{X}_i = \mathbf
{X}(s_i),   \mathbf{Y}_i =
Y(s_i)$, $i=1,\ldots,N $. We construct masked data at $\mathbf{s}$ using
spatial smoothing, and we show later that this masking approach is
a special case of matrix masking by row~(observation)
transformation.

Let $W_\lambda(u,s;\mathbb{S})$ denote the relative weight
assigned to data at location $s$ when generating smoothed data for
the target location $u$, where $\lambda\ge0$ is a smoothness
parameter, and $\mathbb{S}$ denotes all spatial locations in a
study area so $\mathbf{s}$ is a subset of $\mathbb{S}$. The~parameter
$\lambda$ controls the degree of smoothness, with smoothness
increasing with $\lambda$. For notational convenience we suppress
the dependence of $W$ on $\mathbb{S}$.

We consider a subclass of linear smoothers under which the
smoothed spatial processes at location $u$ are defined as
follows. For $\lambda> 0$,
\begin{eqnarray}\label{sm}
Y_{\lambda}(u) & =& \int Y(s)W_\lambda(u,s)\,dN(s) \Big/
\int W_\lambda(u,s)\,dN(s),   \nonumber\\[-8pt]\\[-8pt]
\mathbf{X}_{\lambda}(u) & = & \int\mathbf{X}(s)W_\lambda
(u,s)\,dN(s) \Big/
\int W_\lambda(u,s)\,dN(s)   \nonumber,
\end{eqnarray}
where $N(s)$ is the counting process for locations
with available data from the spatial processes $\{Y(s),\mathbf{X}(s)\}$. For
$\forall u \in\mathbf{s}$ we require that $W_0(u,s) = I_{\{s=u\}}$. If
$W$ is continuous in $\lambda$, we define $W_0(u,s)$ as
$\lim _{\lambda\downarrow0}W_\lambda(u,s)$. Therefore, we
have that $\{Y_{0}(s_i),\mathbf{X}_{0}(s_i)\}=\{Y_i, X_i \}$, the
original individual-level data.

We generate masked data by taking the predictions from~(\ref{sm})
at $\mathbf{s}$ where the original individual-level data are available,
that is, $\{Y_{\lambda}(s_i),\mathbf{X}_{\lambda}(s_i), i=1,\ldots
,N \}$. By
definition in~(\ref{sm}), the masked data are weighted averages of
the original individual-level data $\{Y(s_i),\mathbf{X}(s_i)\}$. The~shape of the weight function $W$ and the degree of smoothness
$\lambda$ control the form and the degree of masking,
respectively, where the degree of masking increases with the
degree of smoothness. In practice, the masked data at location
$s_i$ are computed by
\begin{eqnarray} \label{smsum}
Y_\lambda(s_i) & = &\sum _{k=1}^N Y_k W_\lambda(s_i,s_k)
\bigg/ \sum_{k=1}^N W_\lambda(s_i,s_k), \nonumber \\[-8pt]\\[-8pt]
\mathbf{X}_\lambda(s_i) & = & \sum _{k=1}^N \mathbf{X}_k
W_\lambda(s_i,s_k) \bigg/ \sum_{k=1}^N W_\lambda(s_i,s_k)
,\nonumber
\end{eqnarray}
where the same $\mathbf{W}$ and $\lambda$ are applied to both $Y$
and $\mathbf{X}$. Examples of commonly used smoothers within this class
include parametric linear regressions fitted by ordinary least
square and weighted least square, penalized linear splines with
truncated polynomial basis, kernel smoothers and LOESS
smoothers~[\citet{Simosmoo1996}; \citet{BowmAzzaappl1997}; \citet{HastTibsFrieelem2001}; \citet{ruppert2003}].

Let $\mathcal{Y}$ and $\mathcal{Y}_{\lambda}$ denote the vectors
of $\{Y_i\}$ and $\{Y_{\lambda}(s_i)\}$, and let $\mathcal{X}$ and
$\mathcal{X}_{\lambda}$ denote the matrices of $\{\mathbf{X}_i\}$ and
$\{\mathbf{X}_{\lambda}(s_i)\}$, respectively, where $\mathbf{X}_i$ and
$\mathbf{X}_{\lambda}(s_i),i=1,\ldots, N$, are row vectors. It can
be seen
that $\mathcal{Y}_{\lambda}=\mathcal{A}_\lambda\mathcal{Y}$ and
$\mathcal{X}_{\lambda}=\mathcal{A}_\lambda\mathcal{X}$, where $
\mathcal{A}_\lambda=(\mathcal{A}_{\lambda_{
ij}})=(W_\lambda(s_i,s_j) / \sum_{j=1}^N
W_\lambda(s_i,s_j)  )$. Therefore, constructing
masked data by equation~(\ref{smsum}) is a special case of matrix
masking by row~(observation) transformation. Reidentification from
$(\mathcal{Y}_{\lambda}, \mathcal{X}_{\lambda})$ to $(\mathcal{Y},
\mathcal{X})$ requires knowledge of both $W$ and~$\lambda$ as well
as the existence of $\mathcal{A}_\lambda^{-1}$.

%s2.2 ###
\subsection{Bias and variance in nonlinear regression using masked
data}\label{biasmask}

Bias may arise when a nonlinear model that is specified for the
original individual-level data is fitted to the masked data.
Specifically, we assume the following model for the original
individual-level data which is viewed as the ``truth,'' %
%e2.1 ###
\begin{equation}\label{indimodel}
g(E\{\mathcal{Y}|\mathcal{X}\})= \mathcal{X}\bolds{\beta}.
\end{equation}
Model (\ref{indimodel}) implies the
analogous model for the masked data
%
%e2.2 ###
\begin{equation}\label{maskmodel}
g(E\{\mathcal{Y}_{\lambda}|\mathcal{X}_{\lambda}\})=\mathcal
{X}_{\lambda}\bolds{\beta}
\end{equation}
only for a linear function $g(x)=ax$, where
$a$ is a constant~(except for few special circumstances such as
$\mathbf{X}_{i}={\bf x} $, i.e., constant exposure). Specifically,
\begin{eqnarray*}
g(E\{\mathcal{Y}_{\lambda}|\mathcal{X}_{\lambda}\})&=& a
E\{\mathcal{Y}_{\lambda}|\mathcal{X}_{\lambda}\}= a
E\{\mathcal{Y}_{\lambda}|\mathcal{X}\}\\
&=&a \mathcal{A}_{\lambda}
E\{\mathcal{Y}|\mathcal{X}\}\stackrel{\mathrm{model~(3)}}{=}a \mathcal{A}_{\lambda}a^{-1}\mathcal{X}\bolds{\beta}=
\mathcal{X}_{\lambda}\bolds{\beta}.
\end{eqnarray*}
It follows that for a
nonlinear regression model~(\ref{indimodel}), the coefficient
estimate obtained by fitting model~(\ref{maskmodel}) will be a
biased estimate of $\bolds{\beta}$. Therefore, it is important to evaluate
the bias of the coefficient estimate under model~(\ref{maskmodel})
as well as how the bias varies as a function of the form and the
degree of data masking. To consider both the bias and variance of
the coefficient estimate obtained by fitting
model~(\ref{maskmodel}), we evaluate the MSE as a function of the
form and the degree of masking.

It is common to assume that the masked data are mutually
independent. However, they are generally correlated, since they
combine information across the same original data. To investigate
the impact of this correlation on the uncertainty of the
coefficient estimate when using the masked data, we compare the
``naive'' confidence interval under model~(\ref{maskmodel}) which
does not account for this correlation with an appropriate confidence
interval obtained by using simulation or bootstrap
methods~[\citet{Efroboot1979}; \citet{EfroTibsan1993}].

%s2.3 ###
\subsection{Identification disclosure risk of masked
data}

We evaluate the identification disclosure risk of the masked
data by calculating the probability of identification as developed
in~\citet{Reitesti2005}. To compute the risk of the released
masked data set, we first compute the probability of matching for a
particular data record.

Specifically, let $\mathbf{Z}=(\mathcal{Y},\mathcal{X})$ denote the
unmasked data set and
$\mathbf{Z}_\lambda=(\mathcal{Y}_{\lambda},\mathcal{X}_{\lambda
})$ denote
the released masked data set. Let $\mathbf{t}$ denote a data vector
possessed by a data intruder, where $\mathbf{t}$ contains the true values
for a particular individual. $\mathbf{Z}_\lambda$ can be divided into two
components: $\mathbf{Z}_\lambda^U$ which consists of variables that are
not available\vspace*{-2pt} in $\mathbf{t}$, and $\mathbf{Z}_\lambda^{\mathit{Ap}}$ which
consists of
variables that are available in $\mathbf{t}$. $\mathbf{Z}= (\mathbf
{Z}^U,\mathbf{Z}^{\mathit{Ap}})$ is
the same decomposition of the true data set. Let $J$ be a random
variable that equals $j$ if to match $\mathbf{t}$ with the $j$th
individual in $\mathbf{Z}_\lambda$. The~probability of matching is
$\operatorname{Pr}(J=j|\mathbf{t},\mathbf{Z}_\lambda), j=1,\ldots,N$,
assuming that
$\mathbf{t}$ always corresponds to an individual within~$\mathbf
{Z}_\lambda$.
Assumptions about the knowledge and behavior of the intruder are
used to determine this probability. Using Bayes' rule,
\[
\operatorname{Pr}(J=j|\mathbf{t},\mathbf{Z}_\lambda)=\frac{\operatorname{Pr}(\mathbf{Z}_\lambda
|J=j,\mathbf{t})\operatorname{Pr}(J=j|\mathbf{t})}{\sum_{j=1}^N
\operatorname{Pr}(\mathbf{Z}_\lambda|J=j,\mathbf{t})\operatorname{Pr}(J=j|\mathbf
{t})},
\]
where
$\operatorname{Pr}(\mathbf{Z}_\lambda|J=j,\mathbf{t})$ can be decomposed
into
\begin{eqnarray*}
&&\operatorname{Pr}(\mathbf{z} _{\lambda,1},
\ldots,\mathbf{z} _{\lambda,j-1},\mathbf{z} _{\lambda,j+1},\ldots,\mathbf{z}
_{\lambda,N}
| \mathbf{z} _{\lambda,j}, J=j, \mathbf{t})\\
&&\qquad {}\cdot
\operatorname{Pr}(\mathbf{z} _{\lambda,j}^U|\mathbf{z} _{\lambda,j}^{\mathit{Ap}},J=j,
\mathbf{t}
)\cdot
\operatorname{Pr}(\mathbf{z} _{\lambda,j}^{\mathit{Ap}}|J=j, \mathbf{t}).\nonumber
\end{eqnarray*}
Following
the guidance in~\citet{Reitesti2005}, we compute each component
of $\operatorname{Pr}(J=j|\mathbf{t},\mathbf{Z}_\lambda)$ as follows:
\begin{enumerate}
\item
$\operatorname{Pr}(J=j|\mathbf{t})=1/N$. This is because the true values are
replaced by some weighted averages upon releasing, so exact
matching between $\mathbf{t}$ and any $\mathbf{Z}^{\mathit{Ap}}_{\lambda}$
record is not
possible.
\item$\operatorname{Pr}(\mathbf{z} _{\lambda,j}^{\mathit{Ap}}|J=j, \mathbf{t})$
equals
%
%e2.3 ###
\begin{equation}\label{maskp}
1 - \frac{\Vert \mathbf{z} _{\lambda,j}^{\mathit{Ap}}-\mathbf
{t}\Vert }{\max_{k=1}^N
\Vert \mathbf{z} _{\lambda,k}^{\mathit{Ap}}-\mathbf{t}\Vert },
\end{equation}
which is the tail
probability of a uniform distribution with density $1/\break \max_{k=1}^N
\Vert \mathbf{z} _{\lambda,k}^{\mathit{Ap}}-\mathbf{t}\Vert $. We assume the intruder
knows that
the masked data are weighted averages of the original data. As we
point out at the end of Section~\ref{method}, detailed information
on $W$ and $\lambda$ shall not be released. Therefore, it is a
reasonable assumption that the intruder will assume a uniform
distribution based on the difference from $\mathbf{t}$. The~larger the
difference, the smaller the probability.
\item
$\operatorname{Pr}(\mathbf{z} _{\lambda,j}^U|\mathbf{z} _{\lambda,j}^{\mathit{Ap}},J=j,
\mathbf{t}
)$ is
computed through
\begin{equation}\int
\operatorname{Pr}(\mathbf{z} _{\lambda,j}^U|\mathbf{z} _j^U,\mathbf{z} _{\lambda,j}^{\mathit{Ap}},J=j,
\mathbf{t})\operatorname{Pr}(\mathbf{z} _j^U|\mathbf{z} _{\lambda,j}^{\mathit{Ap}},J=j,
\mathbf{t})\,d
\mathbf{z} _j^U,\nonumber
\end{equation}
where
$\operatorname{Pr}(\mathbf{z} _{\lambda,j}^U|\mathbf{z} _j^U,\mathbf{z} _{\lambda,j}^{\mathit{Ap}},J=j,
\mathbf{t})=1 - \frac{\Vert \mathbf{z} _{\lambda,j}^U-\mathbf{z} _j^U\Vert }{\max_{k=1}^N
\Vert \mathbf{z} _{\lambda,k}^U-\mathbf{z} _j^U\Vert }$,
$\operatorname{Pr}(\mathbf{z} _j^U|\mathbf{z} _{\lambda,j}^{\mathit{Ap}},J=j, \mathbf{t})$
is obtained
through regression of $\mathbf{Z}^U$ on $\mathbf{Z}_\lambda^{\mathit{Ap}}$,
and the
integral is computed using Monte Carlo integration.
\item
$\operatorname{Pr}(\mathbf{z} _{\lambda,1},
\ldots,\mathbf{z} _{\lambda,j-1},\mathbf{z} _{\lambda,j+1},\ldots,\mathbf{z}
_{\lambda,N}
| \mathbf{z} _{\lambda,j}, J=j, \mathbf{t})$ is conservatively assumed
to be
equal to 1. As pointed out in~\citet{Reitesti2005}, such
assumption provides the upper limit on the identification risks and
greatly simplifies the calculation.
\end{enumerate}

Assuming a record $\mathbf{t}$ is matched to the individuals with the
largest probability of matching, we measure the identification
disclosure risk of the entire released data set using the expected
percentage of correct matches. Same as in~\citet{Reitesti2005},
we assume that the intruder possesses correct records for all
individuals in the released data set and seeks to match each record
with an individual with replacement, that is, matching of one record
is independent from matching of another record. Let $m_j$ be the
number of individual records with the maximum matching probability
for $\mathbf{t}_j,j=1,\ldots,N$. Let $I_j=1$ if the $m_j$ individual
records contain the correct match, and $I_j=0$ otherwise. The~expected percentage of correct matches is $\sum_{j=1}^N
\frac{1}{m_j}I_j / N$.

%s3 ###
\section{Simulation studies}\label{sec3}

%s3.1 ###
\subsection{Data generation, parameter estimation and disclosure risk
evaluation}
\label{simuframe}

In this section we conduct simulation studies to illustrate that
parameter estimates from regression using masked data may be less
subject to bias and MSE when the selection of the smoothing weight
function accounts for the spatial patterns of exposure. We
illustrate this point using three examples. In each case, we
define the study area to be $[-1,1]\times[-1,1]$. Within this
study area we randomly select 1000 locations as $\mathbf{s}$ where
individual-level exposure and outcome data are obtained.

In each example, we define a spatial process of exposure $X(s)$
and we obtain $X(s_i)$ for $s_i\in\mathbf{s}$. We simulate the
individual-level outcome data at $\mathbf{s}$ from a model of the general
form
%
%e3.1 ###
\begin{equation}\label{llindi}
Y(s_i)\stackrel{\mathrm{i.i.d.}}{\sim}
\operatorname{Poisson}\bigl(e^{\mu+\beta X(s_i)}\bigr),
\end{equation}
with the individual-level exposure coefficient
$\beta$ being the parameter of interest. The~values of $\mu$ and
$\beta$ are selected to achieve reasonable variability of
$E\{Y(s_i)|X(s_i)\}$
under model~(\ref{llindi}) across the locations. %We then construct
%masked
%data at $\bs$ by spatial smoothing, and we estimate the exposure
%coefficient by fitting the above loglinear model to the masked
%data. We compare such estimate to the true value of $\beta$ to
%explore the resultant bias. In addition, we also compare such
%estimate to that from an analysis using some spatially aggregated
%data, to explore the bias difference between our data masking
%approach and spatial aggregation.

We construct the masked data
$\{Y_{\lambda}(s_i),X_{\lambda}(s_i)\}$ using kernel smoothers,
and we estimate the exposure coefficient $\beta_{\lambda}$ under
model
%
%e3.2 ###
\begin{equation}\label{llmask}
Y_{\lambda}(s_i)\stackrel{\mathrm{i.i.d.}}{\sim}
\operatorname{Poisson}\bigl(e^{\mu_{\lambda}+\beta_{\lambda}
X_{\lambda}(s_i)}\bigr),
\end{equation}
which is analogous to
model~(\ref{llindi}) but fitted to the masked data. The~masked
data are constructed and $\beta_{\lambda}$ is estimated for each
combination of 20 $\lambda$ values and two different kernel
weights, respectively, so we can evaluate the bias and the MSE as
functions of both the smoothing weight and $\lambda$.

In addition, we construct spatially aggregated data by equally
partitioning the\vspace*{1pt} study area into $7\times7=49$ cells and
calculating $Y_{+j}=\sum _{i=1}^{n_j}Y(s_{i})$ and
$\bar{X}_{\cdot j}=\sum _{i=1}^{n_j}X(s_{i})/n_j$, where
$n_j$ is the total number
of individual-level data points in cell~$j$, $j=1,\ldots, 49$. We
estimate the %ecologic
exposure coefficient $\beta_e$ using the aggregated data
$\{Y_{+j},\bar{X}_{\cdot j}\}$ under the analogous model
%
%e3.3 ###
\begin{equation}\label{lleco}
Y_{+j} \stackrel{\mathrm{i.i.d.}}{\sim} n_j\cdot\operatorname{Poisson}\bigl(
e^{\mu_e + \beta_e \bar{X}_{\cdot j}}\bigr).
\end{equation}

To evaluate the identification disclosure risk, we consider the
scenario that a data intruder possesses the correct exposure data,
that is, $X(s_i)$ for $s_i\in\mathbf{s}$, and seeks the matches with the
released data set in order to obtain information on the health
outcome $Y$. Specifically, $\mathbf{Z}^{\mathit{Ap}}$ is $X$ and $\mathbf
{Z}^U$ is $Y$.

We generate 500 replicates of the individual-level outcome data.
For each replicate $\beta_\lambda$ and $\beta_e$ are estimated as
above, and the estimates are averaged across the 500 replicates.

%s3.2 ###
\subsection{Choice of smoothing weight function}

To select a weight function that may lead to less bias and
possibly smaller MSE when estimating the exposure coefficient
using the masked data, we notice that expectation of the masked
outcome $Y_{\lambda}(s_i)$ with respect to model~(\ref{llmask}) is
\[
E\{Y_{\lambda}(s_i)|X_{\lambda}(s_i)\} =
e^{\mu_\lambda+\beta_\lambda
X_{\lambda}(s_i)},
\]
while expectation of
$Y_{\lambda}(s_i)$ with respect to model~(\ref{llindi}) is
\begin{eqnarray*}
E\{Y_{\lambda}(s_i)|\mathbf{X}\} &=&
\int e^{\mu+ \beta X(s)}W_\lambda(s_i, s)\,dN(s)\\
&=&
e^{\mu+ \beta X_{\lambda}(s_i)} \int e^{\beta[X(s)-X_{\lambda
}(s_i)]}W_\lambda(s_i,
s)\,dN(s),
\end{eqnarray*}
where $\mathbf{X}=\{X(s)\}$. The~comparison
between $E\{Y_{\lambda}(s_i)|\mathbf{X}\}$ and\break
$E\{Y_{\lambda}(s_i)|X_{\lambda}(s_i)\}$ suggests that we can
reduce the bias and possibly the MSE of estimating $\mu$ and
$\beta$ when using the masked data by selecting a $W$ s.t. $\int
e^{ \beta[X(s)-X_{\lambda}(s_i)]}W_\lambda(s_i, s)\,dN(s)$ is close
to 1. One way to construct such a $W$ is to assign high weights to
locations that receive similar exposure as the target location and
low weights otherwise. The~$W$ constructed in this way has the
property that it accounts for prior knowledge on the spatial
pattern of the exposure. In our examples, this is also the spatial
pattern of the outcome due to the model assumption~(\ref{llindi}).
Therefore, to assess the difference in bias and MSE when varying
the smoothing weight function, we construct two different kernel
weights for data masking in the way that one weight accounts for
prior knowledge on the spatial pattern of the exposure as above,
while the other does not.

%s3.3 ###
\subsection{Example \textup{I}}

\begin{figure}[b]

\includegraphics{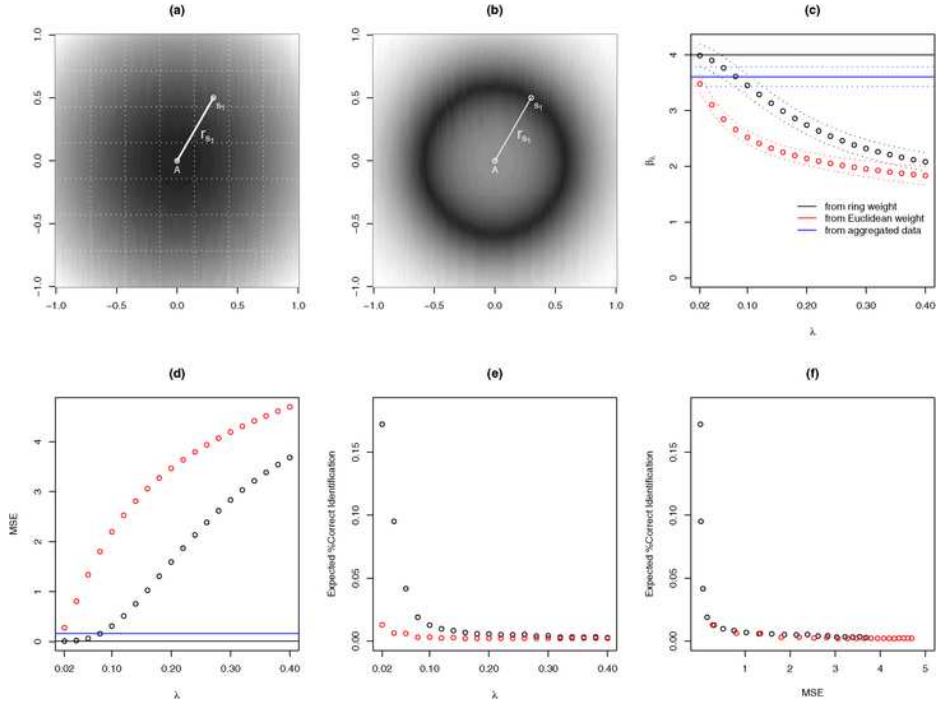}

\caption{Example \textup{I} of
spatially varying exposure, weight function for spatial
smoothing, estimates, and disclosure risk.
 \textup{(a)} Contour plot of exposure from
point source \textup{A}: $X_1(s)=7\exp( -r^2_s / 2.5 )$, with cells for
spatial aggregation.
 \textup{(b)} Contour plot of ring weight function
$W_{1\lambda}(s_1,s) = \exp( -|r^2_s - r^2_{s_1}|
/\lambda)$ for calculating spatially smoothed exposure and
outcome data at location $s_1$, from individual-level exposure
$X_1(s)$ in \textup{(a)} and individual-level outcome $Y_1(s)$ simulated by
$Y_1(s)\sim\operatorname{Poisson}(\exp(-25+ 4 X_1(s)))$ where $\beta=4$,
with $\lambda=0.5$.
 \textup{(c)} Estimates of
$\beta_\lambda$ with ``naive'' 95$\%$ confidence intervals by
fitting model $ Y_{1\lambda}(s)\sim
\operatorname{Poisson}(\exp(\mu_{\lambda}+\beta_{\lambda}
X_{1\lambda}(s)))$ where
$\{Y_{1\lambda}(s),X_{1\lambda}(s)\}$ are constructed using the
ring weight function in \textup{(b)} and using the Euclidean weight
function $W^\ast_\lambda(s_1,s) = \exp( -\Vert s - s_1\Vert ^2 /\lambda
)$, with reference lines at $\beta=4$ and at the estimate from
aggregated data.
 \textup{(d)} Mean square error (MSE) of
$\beta_\lambda$ using ``naive'' variance.  \textup{(e)} Identification disclosure risk measured by the expect
percentage of correct record matching.  \textup{(f)} Disclosure risk versus MSE for utility-risk trade-off.}\label{ringwt}
\end{figure}

We assume that the exposure is eradiated from a point source $A$
and decreases symmetrically in all directions as the Euclidean
distance from $A$ increases. Specifically, we define $
X_1(s)=7\exp( -r^2_s / 2.5 )$ for $s \in[-1,1]\times[-1,1]$,
where $r_s$ is the Euclidean distance between location $s$ and the
point source~$A$. Figure~\ref{ringwt}(a) shows the contour plot
of\break $X_1(s)$. The~individual-level outcome is simulated from
$Y_1(s_i)\stackrel{\mathrm{i.i.d.}}{\sim}\break \operatorname{Poisson}(e^{-25+4%\beta
X_1(s_i)})$. Aggregated data of exposure and outcome are
constructed by calculating group summaries of
$\{Y_1(s_i),X_1(s_i)\}$ as described in Section~\ref{simuframe}.

%f1 ###

We construct masked data $\{Y_{1\lambda}(s_i),X_{1\lambda}(s_i)\}$
by using equation~(\ref{smsum}) with both the Euclidean kernel
weight $W^\ast_\lambda$ and the ring kernel weight $W_{1\lambda}$
which are defined as follows:
%
%e3.5 ###
%e3.4 ###
\begin{eqnarray}
W^\ast_\lambda(u,s) & = & \exp( -\Vert s - u\Vert ^2 /\lambda),\label
{euc}\\
W_{1\lambda}(u,s) & = & \exp( -|r^2_s - r^2_{u}| /\lambda).
\label{ring}
\end{eqnarray}
The~ring kernel weight $W_{1\lambda}(u,s)$
decreases exponentially as the difference between $r^2_s$ and
$r^2_{u}$ increases, and such difference is positively associated
with the difference between $X_1(s)$ and $X_1(u)$ according to the
spatial pattern of the exposure. Figure~\ref{ringwt}(b) shows the
contour plot of $W_{1\lambda}(s_1,\cdot)$. On the other hand, the
Euclidean kernel weight $W^\ast_\lambda(u,s)$ solely depends on
$\Vert s - u\Vert $, the Euclidean distance between location $u$ and
location $s$, and therefore does not account for prior knowledge
on the spatial distribution of the exposure.

%s3.4 ###
\subsection{Example \textup{II}}
\begin{figure}

\includegraphics{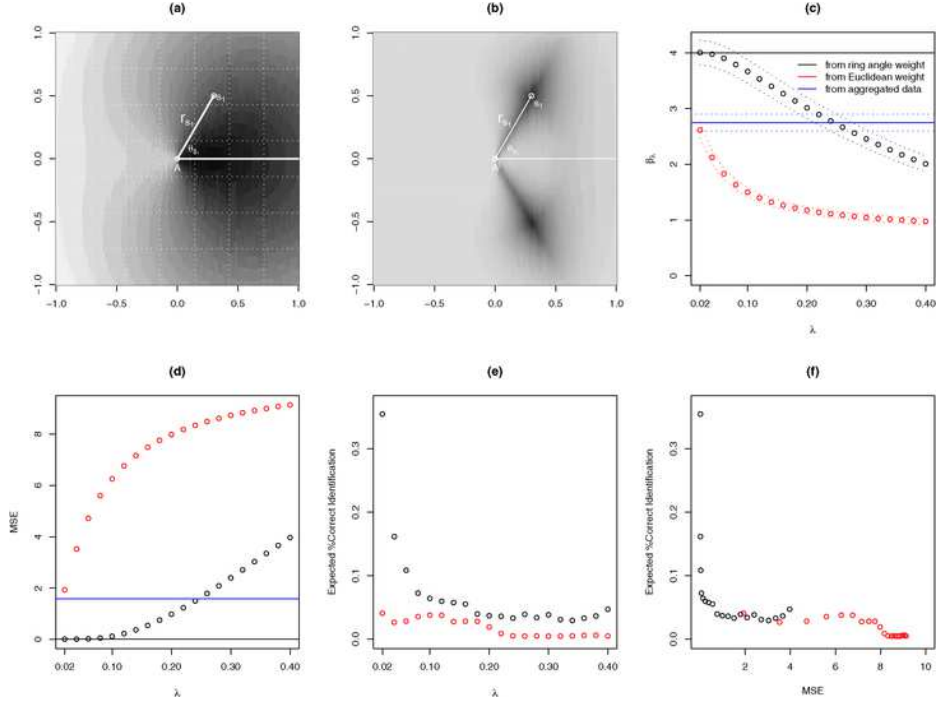}

\caption{Example \textup{II}
of spatially varying exposure, weight function for spatial
smoothing, estimates, and disclosure risk.
 \textup{(a)} Contour plot of exposure from
point source \textup{A} toward a certain direction: $X_2(s)=7\exp(
-r^2_s / 6 - \cos\theta_s / 3)$, with cells for spatial
aggregation.
 \textup{(b)} Contour plot of ring angle weight
function $W_{2\lambda}(s_1,s) = \exp( -(|r^2_s - r^2_{s_1}| +
2|\cos\theta_s - \cos\theta_{s_1}|) /\lambda)$ for calculating
spatially smoothed exposure and outcome data at location $s_1$,
from individual-level exposure $X_2(s)$ in \textup{(a)} and
individual-level outcome $Y_2(s)$ simulated by $Y_2(s)\sim
\operatorname{Poisson}(\exp(-36+\beta X_2(s)))$ where $\beta=4$, with
$\lambda=0.5$.
 \textup{(c)} Estimates of
$\beta_\lambda$ with ``naive'' 95$\%$ confidence intervals by
fitting model
$ Y_{2\lambda}(s)\sim\operatorname{Poisson} (\exp(\mu_{\lambda}+\beta_{\lambda} X_{2\lambda}(s)))$
where $\{Y_{2\lambda}(s),X_{2\lambda}(s)\}$ are constructed using
the ring angle weight function in \textup{(b)} and using the Euclidean
weight function $W^\ast_\lambda(s_1,s) = \exp( -\Vert s - s_1\Vert ^2
/\lambda)$, with reference lines at $\beta=4$ and at the estimate
from aggregated data.  \textup{(d)} Mean square
error~(MSE) of $\beta_\lambda$ using ``naive'' variance.
\textup{(e)} Identification disclosure risk measured by
the expect percentage of correct record matching. \textup{(f)} Disclosure risk versus MSE for utility-risk
trade-off.}\label{ringanglewt}
\end{figure}

We assume that the exposure is eradiated from a point source~$A$
and toward a certain direction. % such as blew by wind.
Specifically, we define $X_2(s)= 7\times \break \exp( -r^2_s / 6 - \cos\theta_s
/ 3)$ for $s \in[-1,1]\times[-1,1]$, where $\theta_s$ is the
angle between the direction from point source $A$ to location $s$
and the direction that the exposure is toward, and $r_s$ is
defined the same as in Example I. Figure~\ref{ringanglewt}(a)
shows the contour plot of $X_2(s)$. The~individual-level outcome
is simulated from $Y_2(s_i)\stackrel{\mathrm{i.i.d.}}{\sim}
\operatorname{Poisson}(e^{-36+%\beta
4X_2(s_i)})$. Aggregated data of exposure and outcome are
constructed by calculating group summaries of
$\{Y_2(s_i),X_2(s_i)\}$ as described in Section~\ref{simuframe}.

We construct masked data $\{Y_{2\lambda}(s_i),X_{2\lambda}(s_i)\}$
by using equation~(\ref{smsum}) with the Euclidean kernel
weight~(\ref{euc}) and the ring angle kernel weight
\[
W_{2\lambda}(u,s) = \exp\bigl( -(|r^2_s - r^2_{u}| + 2 |\cos\theta_s -
\cos\theta_u|) /\lambda\bigr),
\]
which
decreases exponentially as the difference between $r^2_s$ and
$r^2_{u}$ increases as well as the difference between
$\cos\theta_s$ and $\cos\theta_u$ increases.
Figure~\ref{ringanglewt}(b) shows the contour plot of
$W_{2\lambda}(s_1,\cdot)$.

%f2 ###

%s3.5 ###
\subsection{Example \textup{III}}
\begin{figure}

\includegraphics{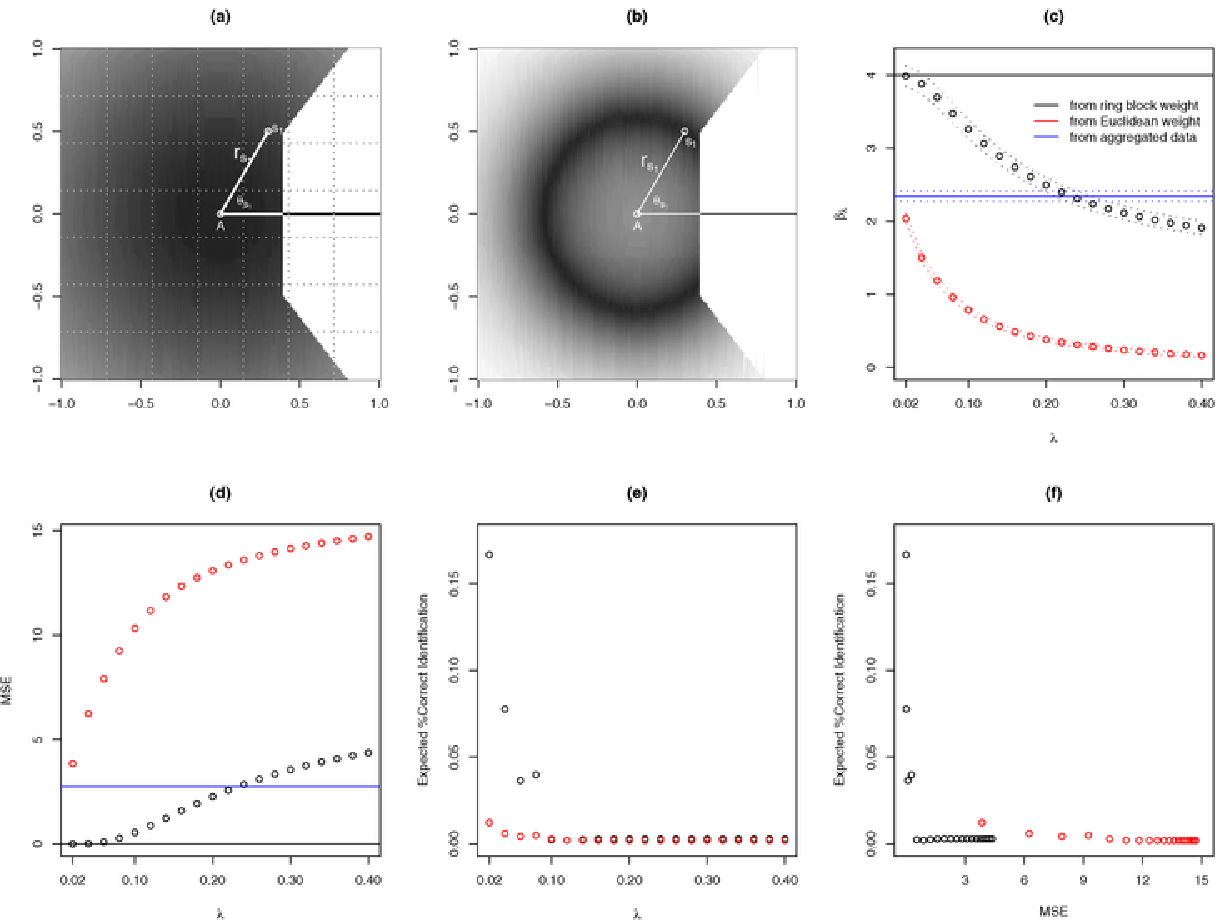}

\caption{Example \textup{III}
of spatially varying exposure, weight function for spatial
smoothing, estimates, and disclosure risk.
 \textup{(a)} Contour plot of exposure from
point source \textup{A} but blocked in certain area: $X_3(s)=7 \exp(
-r^2_s / 2.5 )\cdot I_s$ where $I_s$ is the indicator of location
$s$ in the unblocked area, with cells for spatial aggregation.
 \textup{(b)} Contour plot of ring block weight
function $W_{3\lambda}(s_1,s) = \exp( -|r^2_s - r^2_{s_1}|
/\lambda)\cdot(I_s = I_{s_1})$ for calculating spatially smoothed
exposure and outcome data at location $s_1$, from individual-level
exposure $X_3(s)$ in \textup{(a)} and individual-level outcome $Y_3(s)$
simulated by $Y_3(s)\sim\operatorname{Poisson}(\exp(-24+\beta X_3(s)))$
where $\beta=4$, with $\lambda=0.5$.
 \textup{(c)} Estimates of
$\beta_\lambda$ with ``naive'' 95$\%$ confidence intervals by
fitting model $ Y_{3\lambda}(s)\sim\operatorname{Poisson}(\exp(\mu_{\lambda}+\beta_{\lambda} X_{3\lambda}(s)))$
where $\{Y_{3\lambda}(s),X_{3\lambda}(s)\}$ are constructed using
the ring block weight function in \textup{(b)} and using the Euclidean
weight function $W^\ast_\lambda(s_1,s) = \exp( -\Vert s - s_1\Vert ^2
/\lambda)$, with reference lines at $\beta=4$ and at the estimate
from aggregated data.  \textup{(d)} Mean square
error~(MSE) of~$\beta_\lambda$ using ``naive'' variance.  \textup{(e)} Identification disclosure risk measured by
the expect percentage of correct record matching.  \textup{(f)} Disclosure risk versus MSE for utility-risk
trade-off.}\label{ringblockwt}
\end{figure}

We assume that the exposure is eradiated from a point source $A$
but blocked in a certain area, such as blocked by a mountain, so the
blocked area receives no exposure. Specifically, we define the
unblocked area to be $s_{x} \leq0.4$ or $\cos\vartheta_s
\leq0.625$ for $s \in[-1,1]\times[-1,1]$, where $s_{x}$ is the
$x$-axis value of location $s$ and $\vartheta_s$ is the angle
between the positive $x$-axis and the direction from point source $A$ to
location $s$. We define the exposure $ X_3(s)=7 \exp( -r^2_s / 2.5
) \cdot I_s$ for $s \in[-1,1]\times[-1,1]$, where $I_s$ is the
indicator that $s$ is located within the unblocked area, and $r_s$
is defined the same as in Examples I and II.
Figure~\ref{ringblockwt}(a) shows the contour plot of $X_3(s)$.
The~individual-level outcome is simulated from
$Y_3(s_i)\stackrel{\mathrm{i.i.d.}}{\sim}\operatorname{Poisson}(e^{-24+%\beta
4X_3(s_i)})$. Aggregated data of exposure and outcome are
constructed by calculating group summaries of
$\{Y_3(s_i),X_3(s_i)\}$ as described in Section~\ref{simuframe}.

%f3 ###

We construct masked data $\{Y_{3\lambda}(s_i),X_{3\lambda}(s_i)\}$
by using equation~(\ref{smsum}) with the Euclidean kernel
weight~(\ref{euc}) and the ring block kernel weight
\[
W_{3\lambda}(u,s) = \exp( -|r^2_s - r^2_{u}| /\lambda)\cdot(I_s =
I_u),
\]
which assigns nonzero weight
only when location $u$ and location $s$ are both in the blocked or
unblocked area. In addition, the nonzero weight from
$W_{3\lambda}(u,s)$ decreases exponentially as the difference
between $r^2_s$ and $r^2_{u}$ increases.
Figure~\ref{ringblockwt}(b) shows the contour plot of
$W_{3\lambda}(s_1,\cdot)$.

%s3.6 ###
\subsection{Results}
Results of Example~I on parameter estimates, MSE and
identification risk averaged across the 500 simulation replicates
are shown in Figure~\ref{ringwt}\mbox{(c)--(e)}, respectively.
Specifically, Figure~\ref{ringwt}(c) shows the estimated
$\beta_\lambda$ as a function of $\lambda$ for the ring kernel
weight~(\ref{ring}) and the Euclidean kernel weight~(\ref{euc}),
with the ``naive'' 95$\%$ confidence intervals. By ``naive'' we mean
that the confidence intervals are computed by fitting
model~(\ref{llmask}) directly, and therefore do not account for
the possible correlation between the masked data as pointed out
earlier in Section~\ref{biasmask}. The~reference lines are placed
at the true value of $\beta$ and at the estimated $\beta_e$, from
which the bias of estimating the exposure coefficient by using the
estimated $\beta_\lambda$ can be evaluated.
Figure~\ref{ringwt}(d) shows the MSE as a function of $\lambda$
for the two kernel weights, where in this example MSE is largely
determined by the bias. The~reference lines are placed at the MSE
from regression using the original data~(in which the bias part is
0) and the MSE of $\beta_e$. Figure~\ref{ringwt}(e) shows the
identification disclosure risk of the masked data set measured by
the expected percentage of correct record matching, as a function
of $\lambda$ for the two kernel weights. Figure~\ref{ringwt}(f)
plots the disclosure risk versus MSE, which shows the trade-off
between data utility and disclosure risk.

We find that data masking using the ring kernel
weight~(\ref{ring}) leads to smaller bias and MSE when estimating
the exposure coefficient than masking using the Euclidean kernel
weight~(\ref{euc}), for all $\lambda$ values that are considered.
It suggests that when using the masked data for loglinear
regression, a masking procedure that preserves the spatial pattern
of the original individual-level exposure and outcome data can
lead to better estimates in terms of smaller bias and MSE than a
masking procedure that does not do so. As $\lambda$ increases, the
bias and MSE increase for both kernel weights, while the
differences in the bias and MSE between the two kernel weights
decrease. This increase in the bias/MSE and decrease in the
bias/MSE differences suggest that in the presence of a high degree
of masking, choice for the form of masking may be less influential
on the resultant bias/MSE.
%bias from using the two weight functions will be similar. This
%increase in the bias and decrease in the bias difference between
%the two weight functions is due to the loss of ability for both
%functions to differentiate weights for locations with different
%exposure.\fi
Moreover, comparing the estimated $\beta_\lambda$ and
$\beta_e$, we find that for small values of $\lambda$, the bias
and MSE is smaller when using the estimated $\beta_\lambda$ from
the ring kernel weight~(\ref{ring}).

On the other hand, we find that the disclosure risk is lower when
using the Euclidean kernel weight~(\ref{euc}) for data masking
compared to using the ring kernel weight~(\ref{ring}). This is not
unexpected because masked data constructed using the ring kernel
weight is more informative about the original true values.
However, with a tolerable potential disclosure risk~[$<$0.2 which
is used as an example cutoff in \citet{Reitesti2005}], masked
data when constructed using the ring kernel weight can lead to
better MSE which cannot be achieved by using the Euclidean kernel
weight with a comparable $\lambda$. Same as the trend for bias and
MSE, the differences in the disclosure risk between the two kernel
weights become small as $\lambda$ increases.

Similar results of Example II and Example III are shown in
Figure~\ref{ringanglewt}(c)--(f) and
Figure~\ref{ringblockwt}(c)--(f).

\begin{figure}

\includegraphics{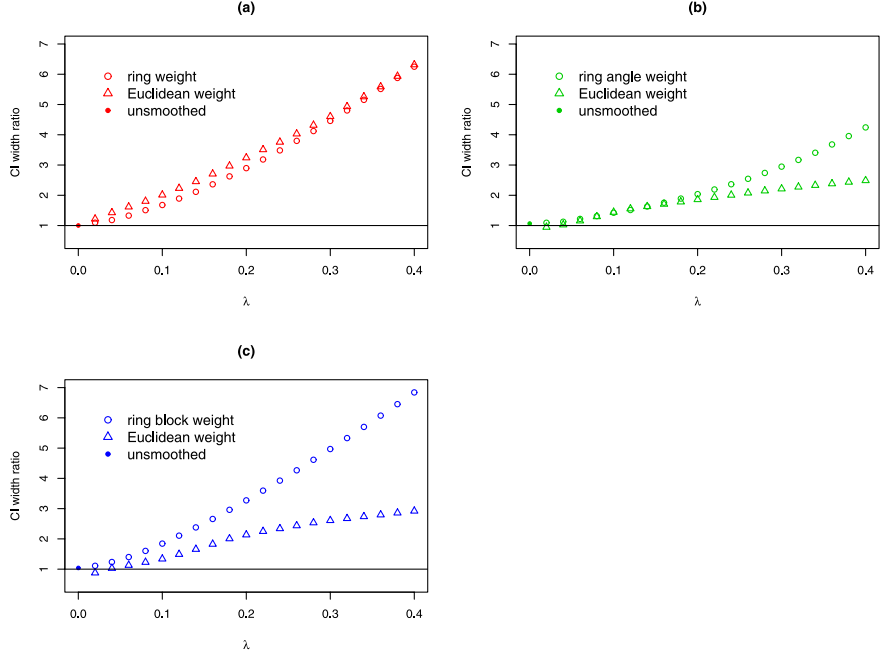}

\caption{Width ratios comparing the 95$\%$ ``naive'' confidence
intervals~(CI) versus the percentile CI obtained from
the empirical distributions of the estimates across the 500
simulations, for the estimates of $\beta_\lambda$ in \textup{(a)}~Example
\textup{I}, \textup{(b)}~Example \textup{II}, and \textup{(c)}~Example \textup{III} of the simulation
studies. Width ratio when $\lambda=0$~(the solid dot) is
calculated using the nonsmoothed
data.}\label{wr}
\end{figure}

Figure~\ref{wr} shows the width ratios comparing the 95\% ``naive'' confidence intervals versus the percentile confidence intervals
obtained from the empirical distribution of the estimates across
the 500 simulations, for the estimates of $\beta_\lambda$ in the
three examples respectively. Width ratio when $\lambda=0$~(the
solid dot) is calculated using the nonsmoothed data, that is, the
individual-level data. We find that in these three examples, the
``naive'' confidence intervals generally overestimate the
uncertainty of the $\beta_\lambda$ estimates, and the degree of
overestimation increases as $\lambda$ increases. In addition, for
Examples II and III where the spatial patterns of exposure are
nonisotropic, the degree of overestimation differs for the weight
functions with and without accounting for prior knowledge on the
spatial pattern of exposure.

%f4 ###

%s4 ###
\section{Application to Medicare data}\label{sec4}
We apply our method to the study of racial disparities in risks of
mortality for a sample of the U.S. Medicare population.

%s4.1 ###
\subsection{Data source}

We extract a large data set at individual-level from the Medicare
government database. Specifically, it includes individual age,
race, gender and a day-specific death indicator over the period
1999--2002, for more than 4 million black and white Medicare
enrollees who are 65 years and older residing in the Northeast
region of the U.S. People who are younger than 65 at enrollment are
eliminated because they are eligible for the Medicare program due
to the presence of either a certain disability or End Stage Renal
Disease and therefore do not represent the general Medicare
population.

\begin{figure}

\includegraphics{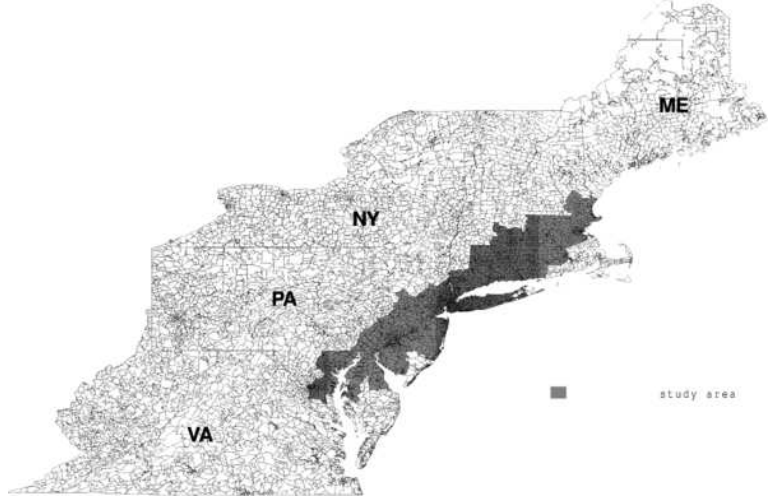}

\caption{Location of the 2095 zip codes
included in our study area.}\label{area}
\end{figure}

Figure~\ref{area} shows the study area which includes 2095 zip
codes in 64 counties in the Northeast region of the U.S. We select the
counties whose centroids are located within the range that covers
the Northeast coast region of the U.S., and we exclude zip codes
without available study population from the study map. This area
covers several large, urban cities including Washington DC,
Baltimore, Philadelphia, New York City, New Haven and Boston. It
has the advantage of high population density and substantial
racial diversity.

%f5 ###

We categorize the age of individuals into 5 intervals based on age
in his/her first year of observation: [65, 70), [70, 75), [75, 80),
[80, 85) and [85, $+$). This categorization facilitates detection
of age effects because differences in the risks of mortality for
one-year increase in age are relatively small. We ``coarsen'' the
daily survival information into yearly survival indicators. By
doing so, we define our outcome as the probability of the occurrence of
death for an individual in one year. This definition adjusts for
the differential follow-up time.

%s4.2 ###
\subsection{Statistical models and data masking}
Let $i$ denote individual, $j$ denote zip code, $t$ denote year,
and $D_{ijt}$ be the death indicator for individual $i$ in zip
code $j$ in year $t$. Similarly as in \citet{zhouracial}, we define
the individual-level model as
\begin{eqnarray}\label{indiindep}
\operatorname{\mathbf{logit}}
\operatorname{Pr}(D_{tij}=1)&=&\beta_0 + \beta_1 \mathit{race}_{ij}
+ \mathit{age}_{ij}\bolds{\beta}_2\nonumber\\[-8pt]\\[-8pt]
&&{} + \beta_3\mathit{gender}_{ij} +
(\mathit{age} \times \mathit{gender}) _{ij}\bolds{\beta}_4.\nonumber
\end{eqnarray}
Geographic locations for each individual
are needed to spatially smooth the individual-level data. However,
from the Medicare data we only have the longitude and latitude of
the zip code centroids. Therefore, we apply a two-step masking
procedure on the individual-level data, where we first aggregate
the individual-level data to zip code-level, and we then spatially
smooth the zip-code level aggregated data to construct the masked
data at the zip code-level.

Specifically, let $D_{++j}$ denote the total death count
and $n_j$ denote the total person-years of zip code $j$. We first
obtain from aggregation $\{\%\ \mathit{black}_{j}$, $\%\ \mathit{agecat}_{j}$,
$\%\
\mathit{male}_{j}$, $\%\ (\mathit{agecat} \times \mathit{male})_{j}$, $p_j=D_{++j}/n_j$, $j=1,\ldots,J\}$, which are the marginal distributions
of race, age, gender, the joint distribution of age and gender,
and the mortality rate, respectively, of each zip code.

Due to the complex spatial pattern of the zip code-level
covariates, we use kernel smoothers with bivariate normal density
kernel weights for spatial smoothing, so the shape of the
smoothing weight is flexible by varying the correlation parameter
value of the bivariate normal distribution. Let the vector
$s=\{s_1,s_2\}$ denote the location of a zip code, where $s_1$ and
$s_2$ are the longitude and latitude of the zip code centroid,
respectively. We use smoothing kernel weights of the general
form
\[
W_\lambda(u,s) = \exp\bigl(-(s_1-u_1,s_2-u_2)^T
\Sigma^{-1}_\lambda(s_1-u_1,s_2-u_2)/2\bigr),
\]
where
\[
 \Sigma_\lambda= \lambda
\pmatrix{
\sigma_1^2 & \rho\sigma_1\sigma_2\vspace*{2pt}\cr
\rho\sigma_1\sigma_2 & \sigma_2^2},
\]
$\sigma_1^2$
and $\sigma_2^2$ are the variances of the longitude and latitude
data of the 2095 zip codes, respectively. %The~R code to calculate
%this weight matrix is provided in \ref{suppA}.
We consider for $\rho$ the following three values:
\begin{enumerate}
\item
$\rho=0$, so the weight solely depends on the Euclidean distance
$\Vert s-u\Vert $;
\item$\rho=0.5$, so higher weight is assigned to $s$
in the northeast and southwest directions of $u$;
\item
$\rho=-0.5$, so higher weight is assigned to $s$ in the northwest
and southeast directions of $u$.
\end{enumerate}

Let $p_{j\lambda}$ denote the smoothed mortality rate of zip code
$j$ from which we calculate the smoothed death count $D_{++j\lambda}=p_{j\lambda}\cdot n_j$. Let $\%\ \mathit{black}_{j\lambda}$,
$\%\  \mathit{agecat}_{j\lambda}$, $\%\  \mathit{male}_{j\lambda}$, $\%\  (\mathit{agecat}
\times \mathit{male})_{j\lambda}$ denote the smoothed marginal
distributions of race, age, gender and the smoothed joint
distribution of age and gender, respectively, of zip code $j$. We
specify the model for masked data as
%
%e4.1 ###
\begin{eqnarray}
\label{ecoindep}
D_{++j\lambda} & \sim&
\operatorname{Bin}(n_j,p_{j\lambda}),
\nonumber\\
\operatorname{logit} p_{j\lambda} & = & \beta_{0\lambda} +
\beta_{1\lambda}\%\ \mathit{black}_{j\lambda} +
\bolds{\beta}_{2\lambda}\%\
\mathit{agecat}_{j\lambda}\\
&&{} + \beta_{3\lambda}\%\  \mathit{male}_{j\lambda} +
\bolds{\beta}_{4\lambda}\%\  (\mathit{agecat} \times \mathit{male})_{j\lambda}.
\nonumber
\end{eqnarray}
The~zip code-level nonsmoothed aggregated data are also used to fit
model (\ref{ecoindep}).

To evaluate the identification disclosure risk, we consider the
scenario that a data intruder possesses correct zip code-level
demographic data and seeks the matching with the masked zip
code-level data set in order to obtain information on the zip
code-level mortality. Specifically, the released data set consists
of $\%\ \mathit{black}_{j\lambda}$, $\%\  \mathit{agecat}_{j\lambda}$,
$\%\
\mathit{male}_{j\lambda}$ and $p_{j\lambda}, j=1,\ldots,2095$, and the data
intruder possess the correct $\%\ \mathit{black}_{j}$, $\%\  \mathit{agecat}_{j}$ and
$\%\  \mathit{male}_{j}$.

%s4.3 ###
\subsection{Choice of association measure}
The~common approach to report the association between race and
mortality risks is to report the race coefficients $\beta_{1}$ in
model~(\ref{indiindep}) and $\beta_{1\lambda}$ in
model~(\ref{ecoindep}), whose interpretation is subjected to the
coding of the race covariate. For direct understanding of the
difference in the risk of death between the black and white
populations, we define and report the population-level odds
ratio~(OR) of death comparing Blacks versus Whites, which is a
function of the predicted values [\citet{zhouracial}]. Therefore,
interpretation of this association measure does not depend on
model parameterization (e.g., on covariate centering and scaling).

Specifically, let
\begin{eqnarray*}
P_{tijb}&=&\operatorname{Pr}(D_{tij}=1|
\mathit{race}_{ij}=\mathit{Black}, \mathit{age}_{ij},
\mathit{gender}_{ij}),\\
P_{tijw}&=&\operatorname{Pr}(D_{tij}=1|\mathit{race}_{ij}=\mathit{White},\mathit{age}_{ij},
\mathit{gender}_{ij})
\end{eqnarray*}
denote the predicted probabilities of death in
year $t$ for a black person and a white person, respectively,
whose other covariates values are the same as the $i$th individual
in the $j$th zip code. We define the population-level OR from the
individual-level model~(\ref{indiindep}) as follows:
\[
\mathit{OR} =
\frac{P_{\cdot\cdot\cdot b}Q_{\cdot\cdot\cdot w}}{P_{\cdot
\cdot\cdot w}Q_{\cdot\cdot\cdot
b}},
\]
where
\begin{eqnarray*}
P_{\cdot\cdot
\cdot b} &=&
\sum_{t,i,j} P_{tijb},\qquad P_{\cdot\cdot\cdot w} = \sum_{t,i,j}
P_{tijw},\\
 Q_{\cdot\cdot\cdot b} &=& 1 - P_{\cdot\cdot\cdot
b},\qquad Q_{\cdot\cdot\cdot w} = 1 - P_{\cdot\cdot\cdot
w}.
\end{eqnarray*}
Similarly, we define population-level $\mathit{OR}_\lambda$ from
model~(\ref{ecoindep}) using summary probabilities
\[
P_{\cdot
b\lambda} = \frac{\sum_j n_j P_{jb\lambda}}{\sum_j
n_j}\quad \mbox{and}\quad  P_{\cdot w\lambda} = \frac{\sum_j n_j
P_{jw\lambda}}{\sum_j n_j},\nonumber
\]
where $P_{jb\lambda}$ and
$P_{jw\lambda}$ are the predicted probabilities of death in one
year for zip codes that consist of solely black and solely white
populations, respectively, and whose marginal and joint
distributions of age and gender are the same as zip code~$j$.
``Naive'' standard errors of $\log\mathit{OR}_\lambda$ are calculated
using the multivariate Delta Method~[\citet{CaseBergsta2002}].
In addition, bootstrap confidence intervals for $\log\mathit{OR}_\lambda$
are calculated using 1000 nonparametric bootstrap samples. Both
``naive'' and bootstrap confidence intervals for $\mathit{OR}_\lambda$ are
obtained by exponentiating the corresponding confidence intervals
for $\log\mathit{OR}_\lambda$.

%s4.4 ###
\subsection{Results}\label{darlt}

\begin{figure}

\includegraphics{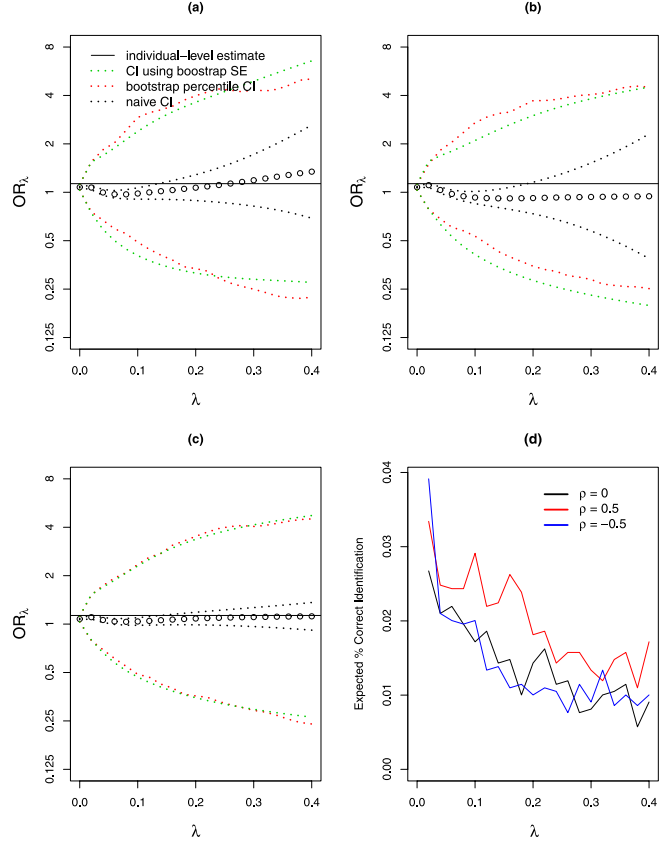}

\caption{Estimates of
$\mathit{OR}_\lambda$ under model \textup{(\protect\ref{ecoindep})} and identification
disclosure risk as a function of $\lambda$ for the three weight
functions. Estimates of $\mathit{OR}_\lambda$ is plotted with the 95$\%$
``naive'' confidence intervals (CI), CI using bootstrap standard
error (SE) estimates, and bootstrap percentile CI. $\mathit{OR}_{0}$ is
estimated by fitting model \textup{(\protect\ref{ecoindep})} to the nonsmoothed
zip code-level aggregated data.
 \textup{(a)} Estimates of $\mathit{OR}_\lambda$ for
bivariate normal density kernel weight with $\rho=0$.
 \textup{(b)} Estimates of $\mathit{OR}_\lambda$ for
bivariate normal density kernel weight with $\rho=0.5$.
 \textup{(c)} Estimates of $\mathit{OR}_\lambda$ for
bivariate normal density kernel weight with $\rho=-0.5$.
 \textup{(d)} Identification disclosure risk
measured by expected percentage of correct matching.}\label{ORsehalf}
\end{figure}

Figure~\ref{ORsehalf}(a)--(c) shows the estimates of
$\mathit{OR}_\lambda$ under model~(\ref{ecoindep}) as a function of
$\lambda$ for the three kernel weights respectively, with the
95\% ``naive'' confidence intervals, confidence intervals using
bootstrap standard error estimates and bootstrap percentile
confidence intervals. $\mathit{OR}_{0}$ is estimated by fitting
model~(\ref{ecoindep}) to the nonsmoothed zip code-level
aggregated data. The~reference line is placed at the estimate of
OR under the individual-level model~(\ref{indiindep}).

%f6 ###

For small values of $\lambda$ ($<$0.1), the estimates of
$\mathit{OR}_\lambda$ for all three kernel weights are smaller than the
estimate of OR and therefore produce negative bias, while for
larger values of $\lambda$ the bias differs substantially for
different kernel weights. For example, data masking using the
kernel weight with $\rho=0.5$ leads to consistent underestimation of
the odds ratio for all $\lambda$ values that are considered. When
using the kernel weight with $\rho=-0.5$ for data masking, the
estimates of $\mathit{OR}_\lambda$ are less subject to bias than those from
using the other two kernel weights. Differences in MSE between the
three kernels can also be inferred from the plots, and we find
that using the kernel weight with $\rho=-0.5$ leads to much
smaller MSE than using the other two kernels. For all three kernel
weights, the ``naive'' confidence intervals underestimate the
uncertainty of the $\mathit{OR}_\lambda$ estimates, which is in the
opposite direction of the relation between the ``naive'' and the
appropriate confidence intervals in the simulation studies. The~two bootstrap confidence intervals are wider than the ``naive'' confidence interval when $\lambda=0$, which suggests a systematic
difference between the bootstrap confidence intervals and the
``naive'' confidence intervals regardless of smoothing. This
systematic difference occurs because the nonsmoothed zip
code-level aggregated data may not satisfy the Binomial model
assumption in~(\ref{ecoindep}).

Figure~\ref{ORsehalf}(d) shows the identification disclosure risk
of the masked data set as measured by the expected percentage of
correct matches when using the three kernel weights for masking,
as a function of $\lambda$. The~disclosure risk for all three
kernel weights are small, ranging from 0.01--0.04. The~risk is
similar for the masked data sets when using the kernel weight with
$\rho=0$ and $\rho=-0.5$ for masking, and the risk when $\rho=0.5$
is slightly higher.

%s5 ###
\section{Discussion}\label{sec5}
We propose a special case of matrix masking based on spatial
smoothing techniques, where the smoothing weight function controls
the form of masking, and the smoothness parameter value directly
measures the degree of masking. Therefore, data utility and
disclosure risk can be calculated as functions of both the form
and the degree of masking. In fact, the smoothing weight function~$W$ can be any weight function and is not restricted by existing
smoothing methods. With the variety of combinations of weight
functions and smoothness parameter values, it is feasible to
construct masked data that maintain high data utility while
preserving confidentiality.

We consider a subclass of linear smoothers that produces masked
data as weighted averages of the original data. Therefore, the
masked data values are within a reasonable range. More
importantly, correlation among the variables is invariant under
linear transformation, which may intrinsically contribute to
better data utility of the masked data. On the other hand, this
subclass is a large class. It includes many commonly used
smoothers. We do not expect major restriction by focusing on this
subclass of linear smoothers.

Using our method, we investigate the utility of the masked data in
terms of bias, variance and MSE of parameter estimates when using
the masked data for loglinear and logistic regression analysis.
Note that similar studies can be applied to any GLM. In addition,
we evaluate the identification disclosure risk of the masked
data set by calculating the expected percentage of correct record
matching. In the simulation studies, we provide guidance for
constructing masked data that can lead to better regression
parameter estimates in terms of smaller bias and MSE for loglinear
models, and we show the trade-off between better estimates and
lower disclosure risk. Specifically, masked data can be
constructed by using a smoothing weight function that accounts for
prior knowledge on the spatial pattern of individual-level
exposure, together with a reasonably low degree of masking. We
provide guidance for how to select such a smoothing weight
function for loglinear models. In addition, we provide candidate
weight functions for three simplified but representative spatial
patterns of exposure.

As is expected, masked data that can lead to better estimates are
generally more informative about the original data values and
therefore are subject to relatively higher identification
disclosure risk. However, the flexibility in our data masking
method enables constructing the masked data that can lead to good
parameter estimates, while the disclosure risk is controlled at a
low level. In the meanwhile, caution should be placed to the
institute in releasing detailed information on the data masking
approach along with masked data. It is pointed out in
Section~\ref{method} that simultaneously releasing the smoothing
weight function $W$ and the smoothness parameter $\lambda$ in the
existence of $\mathcal{A}_\lambda^{-1}$ can lead to
reidentification of original data. However, even if only partial
information is released, for example, only the information that
data are masked using smoothing and the smoothing weight function
is released while the smoothing parameter value is not released,
it is possible that a smart data intruder can still reconstruct
the transformation matrix $\mathcal{A}_\lambda$.

We apply our data masking method to the study of racial
disparities in risks of mortality for the Medicare population, and
show how the bias and the variance of the estimated OR of death
comparing blacks to whites, and how the identification disclosure
risk, vary with the form and the degree of masking. The~results
suggest that in the absence of clear guidance, it is helpful to
explore a large flexible family such as the bivariate normal
density kernel to identify a weight function that can lead to both
good utility and low identification risk for the masked data.

We compare the ``naive'' confidence intervals with the appropriate ones which account for
the possible correlation among masked data in both the simulation
studies and the data application, where we observe opposite
directions in the relation between the ``naive'' and the
appropriate confidence intervals. It suggests no general direction
for that relation. One possible reason, which is also pointed out
in Section~\ref{darlt}, is that the unmasked data in the
simulation study are simulated from Poisson distributions, while
the unmasked data in the data application are real data and do not
strictly follow the assumed binomial distribution. Therefore, in
the data application, the standard errors account for both the
correlation among the masked data and the discrepancy of the
original data distribution from binomial.

The~simulation study and data application results show that masked
data constructed using our method can well preserve
confidentiality. Specifically, the identification disclosure risk
is reasonably low for all scenarios that we consider. Note that
our calculation of the disclosure risk is conservative: we assume
that an intruder possesses true values for all the regressors, and
we use probability 1 for the component $\operatorname{Pr}(\mathbf{z} _{\lambda,1},
\ldots,\mathbf{z} _{\lambda,j-1},\mathbf{z} _{\lambda,j+1},\ldots,\mathbf{z}
_{\lambda,N}
| \mathbf{z} _{\lambda,j},\break J=j, \mathbf{t})$ in the calculation. In
addition, the
flexibility in the selection of smoothing weight function $W$ and
smoothness parameter $\lambda$ can also help control disclosure
risk in addition to improving data utility.

Based on our method, we additionally derive a closed-form
expression for first-order bias of the parameter estimates
obtained using the masked data, for GLM that belong to the
exponential family. The~first-order bias calculation is not
necessary when both individual-level exposure and health outcome
data are available so the actual bias can be computed. It may be
used by researchers who have only the individual-level exposure
information to explore the possible bias in their analysis using
masked data.

Although our proposed method uses spatial smoothing and therefore
applies to spatial data, it can be easily generalized to other
data types because the masking procedure is a smoothing technique
that takes weighted averages of the original data. For example,
the proposed method can be generalized to smoothing time series
data by using the smoothing weight function $W_\lambda(\mu,s)$, where
$\mu$ and $s$ denote time points. Also, note that an alternative
method to mask spatial data is to mask the individual spatial
location [see \citet{ArmsRushZimmgeog1999}; \citet{wieland2008}].

\begin{appendix}\label{append}

\section*{Appendix: First-order bias}\label{fobpure}
We derive a closed-form
expression for the first-order
bias of estimating the regression coefficients in a GLM that belongs
to the exponential family, when using data masked by our method.
Let $\bolds{\beta}$ denote the vector of regression coefficients of a
model specified for the original individual-level data. When the
model belongs to the exponential family, its log likelihood can be
expressed as
\[
\operatorname{LL}(\bolds{\beta}) = \sum_{i=1}^N\frac{Y_i \mathbf
{X}_i\bolds{\beta}-
b(\mathbf{X}_i\bolds{\beta})}{a(\phi)} + C(Y_i,\phi),
\]
$b^\prime(\mathbf{X}_i\bolds{\beta})=g^{-1}(\mathbf{X}_i\bolds
{\beta})$, where $b^\prime(\cdot)$
is the derivative of function $b(\cdot)$, and $g(\cdot)$ is the
link function. Substituting the individual-level data $\{Y_i,
\mathbf{X}_i\}$ by the masked data $\{Y_{\lambda}(s_i),
\mathbf{X}_{\lambda}(s_i)\}$, we obtain log likelihood of the analogous
model when fitted to the masked data,
%
%e5.1 ###
\begin{eqnarray}\label{llsm}
\qquad \operatorname{LL}_{\mathrm{m}}(\bolds{\beta}_\lambda;\lambda)&=&\sum_{i=1}^N\frac
{Y_{\lambda}(s_i)
\mathbf{X}_{\lambda}(s_i)\bolds{\beta}_\lambda-
b(\mathbf{X}_{\lambda}(s_i)\bolds{\beta}_\lambda)}{a_\lambda(\phi
_\lambda)} +
C_\lambda(Y_{\lambda}(s_i),\phi_\lambda),
\end{eqnarray}
where
$\bolds{\beta}_\lambda$ denotes the corresponding vector of regression
coefficients. In order to calculate the MLE of $\bolds{\beta}_\lambda
$, it
is common procedure to calculate the score function from the
likelihood (\ref{llsm}) and take its expectation with respect to
the ``true'' individual-level model $E\{Y_i|\mathbf{X}_i\}$. Denote the
expected score function as $\overline{S}(\lambda,\bolds{\beta
}_\lambda)$
and denote $\bolds{\beta}(\lambda)$ as the solution s.t.
$\overline{S}(\lambda,\bolds{\beta}(\lambda))=0$. It can be shown that
$\bolds{\beta}(0)=\bolds{\beta}$. Taking the derivative of
$\overline{S}(\lambda,\bolds{\beta}(\lambda))=0$ with respect to
$\lambda$
and evaluating it at $\lambda=0$, we obtain the standard result:
%
%e5.2 ###
\begin{equation}\label{betaprime}
\bolds{\beta}^{\prime}(0) = -
(\overline{S}_2(0,
\bolds{\beta}(0) )^{-1}\cdot\overline{S}_1(0, \bolds{\beta}(0)) ,
\end{equation}
where
$\overline{S}_1$ and $\overline{S}_2$ are the partial derivatives
with respect to the first and second components of
$\partial\overline{S}/\partial\lambda$, respectively.
Specifically,
%
%e5.3 ###
\begin{eqnarray}\label{FOB}
\overline{S}_1(0, \bolds{\beta}(0)) & = &
\sum_{i=1}^N  \mathbf{X}_i^T \biggl(\int h(\mathbf{X}(s)\bolds
{\beta})R_0(s_i, s)\,dN(s)\nonumber\\
&&\hspace*{35pt}{} -
h^{\prime}(\mathbf{X}_i\bolds{\beta}) \int\mathbf{X}(s)^T
R_0(s_i, s)\,dN(s)\cdot\bolds{\beta}
\biggr) \\
\overline{S}_2(0, \bolds{\beta}(0)) & = & - \sum_{i=1}^N~ h^{\prime
}(\mathbf{X}
_i\bolds{\beta})\cdot\mathbf{X}_i^T\mathbf{X}_i,\nonumber
\end{eqnarray}
where $R_0(s_i, s)=  \frac{\partial (W_\lambda(s_i, s)/\int
W_\lambda(s_i,s)\,dN(s))}{\partial\lambda}|_{\lambda=0}$ and
$h(\cdot)=g^{-1}(\cdot)$, inverse of the link function of the
GLM. In practice, $\overline{S}_1(0, \bolds{\beta}(0))$ in~(\ref
{FOB}) is
calculated by substituting the the integrals by summations over
all locations where the original individual-level data are
available.

The~quantity $ \bolds{\beta}^{\prime}(0)$ denotes the instant bias of
estimating $\bolds{\beta}$ %that results from fitting GLM
using masked data, when changing from no masking to a very low
degree of masking. As expected, when (i) $\mathbf{X}(s)$ is constant
across all locations in $\mathbf{s}$, or (ii) $g(\cdot)$ is a linear
function, $\overline{S}_1(0, \bolds{\beta}(0))$ is calculated to be
0, and
therefore $\bolds{\beta}^{\prime}(0)=0$.

Using $\bolds{\beta}^\prime(0)$, we can approximate the bias of estimating
$\bolds{\beta}$ when fitting a GLM using masked data whose degree of masking
is $\lambda$, by calculating
\[
\bolds{\beta}(\lambda)-\bolds{\beta}\approx
\bolds{\beta}^{\prime}(0)\cdot\lambda.
\]
This bias calculation
can be extended to any function of $\bolds{\beta}$, for example, the
predicted value. Specifically, bias in estimating $f(\bolds{\beta})$ can
be approximated by
\[
f(\bolds{\beta}(\lambda))-f(\bolds{\beta})\approx
f^\prime(\bolds{\beta})\cdot\bigl(\bolds{\beta}(\lambda)-\bolds
{\beta}\bigr)\approx
f^\prime(\bolds{\beta})\cdot\bolds{\beta}^{\prime}(0)\cdot
\lambda.
\]

It can be seen that the first-order bias approximation can be
easily generalized to approximation using higher-order terms of
the Taylor series expansion in addition to the first-order term.
Specifically,
\begin{eqnarray}\label{taylor}
\bolds{\beta}(\lambda)-\bolds{\beta}&\approx&
\bolds{\beta}^{\prime}(0)\cdot\lambda+
\bolds{\beta}^{\prime\prime}(0)\cdot\lambda^2/2 + \cdots\nonumber\\[-8pt]\\[-8pt]
&&{}+
\bolds{\beta}^{(n)}(0)\cdot\lambda^n/n!,\qquad n \geq1.\nonumber
\end{eqnarray}
Similarly, we can generalize the bias approximation of estimating
$f(\bolds{\beta})$.

A limitation of the bias approximation using Taylor series
expansion~(\ref{taylor}) is that we ignore the remainder term
$\bolds{\beta}^{(n+1)}(\xi)\cdot\frac{\lambda^{n+1}}{(n+1)!}$,
$\xi\in(0,\lambda)$, which may not be\vspace*{1pt} small for large values of
$\lambda$. Therefore, the approximation only captures the bias for
$\lambda\approx0$, that is, the instant direction and magnitude of
the bias when changing from no masking to a very low degree of
masking. It may not capture the total bias for a specified degree
of masking. In the application of our method to the Medicare data,
the first-order bias is calculated to be 0 for all three kernel
weights because $R_0$ in~(\ref{FOB}) equals 0. In addition, when
applying the bias approximation~(\ref{taylor}) to the three
examples in the simulation studies for $n=1,\ldots,5$, the bias
approximation is calculated to be 0, while nonzero bias is shown
by comparing the parameter estimates when using the masked data
with the true parameter value.
\end{appendix}

\section*{Acknowledgment}
Thanks to Dr. Aidan McDermott for
the help on the Medicare data sources.

\begin{supplement}[id=suppA]
\sname{Supplement}
\stitle{R code}
\slink[doi]{10.1214/09-AOAS325SUPP}
\slink[url]{http://lib.stat.cmu.edu/aoas/325/supplement.zip}
\sdescription{We provide the R code for (1) the simulation study
utility part of the three examples,
(2) the function to compute the disclosure risk, and (3) the
calculation of the bivariate normal density kernel weight matrix.}
\end{supplement}

\printaddresses

\end{document}